\documentclass[11pt]{article}
\usepackage{amssymb}
\usepackage{graphics,epsfig}
\usepackage{here}
\usepackage{bdrep}
\usepackage{bdpage}
\TextSize{24.5cm}{14cm}

\let\text\mbox
\let\epsfig\psfig
\newtheorem{remarkt}{Remark}

\begin{document}

\ReportSmall
{
	Fairness issues in a chain of IEEE 802.11 stations
}
{
\begin{tabular}{lll}
Bertrand Ducourthial & Yacine Khaled & St\'ephane Mottelet\\
\multicolumn{2}{c}{Heudiasyc lab., UMR-CNRS 6599} & LMAC lab., EA 2222\\
\end{tabular}

Universit\'e de	Technologie de Compi\`egne\\
B.P. 20529, F-60205 Compi\`egne cedex, FRANCE\\ 
Email: firstname.name@utc.fr
}
{
We study a simple general scenario of ad hoc networks based on IEEE
802.11 wireless communications, consisting in a chain of transmitters,
each of them being in the carrier sense area of its neighbors. Each
transmitter always attempts to send some data frames to one receiver in
its transmission area, forming a pair sender-receiver. This scenario
includes the three pairs fairness problem introduced in
\cite{Dhoutaut02}, and allows to study some fairness issues of the IEEE
802.11 medium access mechanism.

We show by simulation that interesting phenomena appear, depending on
the number $n$ of pairs in the chain and of its parity. We also point
out a notable asymptotic behavior.
We introduce a powerful modeling, by simply considering the probability
for a transmitter to send data while its neighbors are waiting. This
model leads to a non-linear system of equations, which matches very well
the simulations, and which allows to study both small and very large
chains. We then analyze the fairness issue in the chain regarding some
parameters, as well as the asymptotic behavior.
By studying very long chains, we notice good asymptotic fairness of the
IEEE 802.11 medium sharing mechanism.
As an application, we show how to increase the fairness in a chain of
three pairs.
}
{%
July 2005
}

\tableofcontents

\newpage

\section{Introduction}
\label{s:introduction}

\subsection{Motivations}
Recently, wireless networks have increasingly received attention
from the networking community. Although several wireless communication
standards have been proposed, the IEEE 802.11 protocol
\cite{IEEE802.11,IEEE802.11b,IEEE802.11a} is the most widely used, and
constitutes the de facto solution for practical network connection
offering mobility, flexibility, low cost of deployment and use.
This success leads to many studies of the protocol, in various
situations (either ad hoc or with access point) and by different means
(experimentation, simulation, modeling).
It remains that, besides its qualities, the 802.11 protocol, and
particularly its medium access control mechanism, suffers from some
imperfections in terms of global throughput and fairness between nodes.
Our work deals with some fairness issues with 802.11 protocol in ad hoc
mode.

\begin{figure}[b!]
	\centering
	\includegraphics[width=\linewidth]{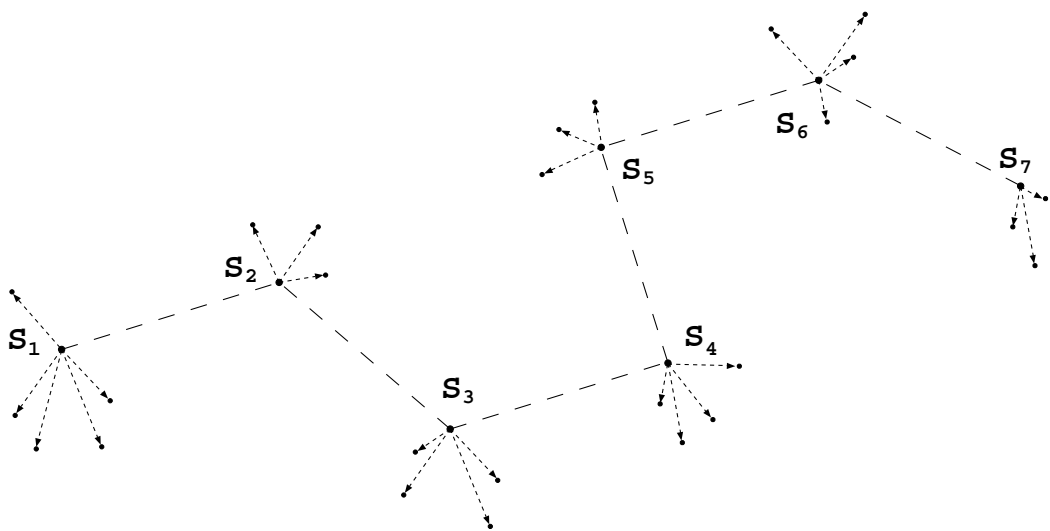}
	\caption{A chain of \emph{senders}.\label{f:chain}}
\end{figure}

We study a simple but general scenario, where some nodes (hereby called
\emph{senders}) try to continuously send some data to one of their
neighbors (hereby called \emph{receiver}), not necessarily always the
same. The senders form a chain, each of which being in the carrier sense
area of its neighbors (Figure~\ref{f:chain}).

In \cite{Dhoutaut02}, the authors study a similar scenario composed of
three pairs, and shows that the central pair obtains a very poor
throughput compared to the border pairs. For instance, with a sending
rate of 2\,Mbits/s, the central pair has only a throughput of
0.04\,Mbits/s compared to 1.55\,Mbits/s for the external ones (the
throughput of a single alone pair is 1.59\,Mbits/s in this situation).

This scenario is a particular case of the chain of senders scenario we
study in this paper. It combines both EIFS delay mechanism and asymmetry
of the chain in terms of number of neighbor senders.  We show that
interesting phenomena appear when the number of pairs increases in the
chain. These phenomena depend on the number $n$ of pairs as well as on
its parity. Moreover a notable asymptotic behavior appears when $n$
increases. We provide a powerful modeling which leads, among others, to
interesting conclusions in terms of fairness both for small and large
chains. This analysis allows us to better understand the DCF properties
and to improve the fairness in a chain, especially in the three pairs
case.

\subsection{Related work} 

There is a large amount of literature dealing with the performances of
the IEEE 802.11 \emph{Distributed Coordinated Function} (DCF)
responsible of the shared radio medium sharing.


In \cite{Cali98}, a relation between the necessary and real time for
sending some data is given, allowing to estimate the DCF capacity.
%
In \cite{Bianci00} the authors make an analytical study of the rates
calculation of the DCF using Markov chain.
The authors prove that the performances of the DCF depends on the
minimal contention window and on the number of stations in the network.

In \cite{Heindl01}, a modeling of the IEEE 802.11 DCF with stochastic
Petri nets is proposed. Among other results, the authors show that the
EIFS delay used when a collision occurs can be advantageous when the
network is not saturated.



In \cite{Vishnevsky02a}, the authors modify the model suggested in
\cite{Bianci00}, and give an estimation of the throughput as a function
of the number of stations in the network and of the ambient noise.
Reusing works of \cite{Bianci00, Cali98}, the authors improve their
results in \cite{Vishnevsky02b}, by taking into account the contention
window increasing in case of collision.

Besides throughput evaluation, some studies deal with the DCF fairness.

In \cite{Bharghavan94}, the authors present a case where the binary
exponential backoff (BEB) lead to an unfair situation.
Indeed, consider a situation where the contention window of the
competing transmitters are large due to collisions. As soon as a node
succeeds in sending a frame, it will reset its contention window. As a
consequence, it will generally wait for smaller backoff than others for its further
transmissions, and then gain more easily access to the shared medium.
To resolve such problems, the authors design the medium access protocol
MACAW.



In \cite{Li04b}, the authors present the relevance of the EIFS mechanism
to the fairness.
They show that the EIFS delay can be too large or too small according to
some scenarios.  The authors propose then an adaptive mechanism for
determining the EIFS delay, based on a measurement of the occupation
time of the medium.

In \cite{Nandagopal00} the authors propose an evaluation of the DCF
fairness, by means of maximization of some differentiable concave
functions, under a set of constraints representing the impossibility for
two close transmitters to simultaneously transmit a frame with success.
Some unfair situations relying on asymmetric topologies are studied by
simulation. They also study fairness per packets and fairness per flow:
two mobiles with the same probability of access to the medium do not
constitute an equitable scenario when one of both must retransmit more
flow than the other.



In \cite{Dhoutaut02,Chaudet05}, the authors study an unfair scenario called
\emph{three pairs problem} by means of simulations and
experimentations. This scenario relies on an asymmetric topology
composed of three pairs of nodes. 
Pair 2 is placed between pairs 1 and 3, and is in the carrier sense of
its both neighbors. The emissions of pairs 1 and 3 are not synchronized,
and when the pair 2 wants to emit, it is necessary that the silence
periods of the other mobiles overlap. However the probability of such a
covering is weak.

This scenario has been modeled in \cite{Chaudet04b} with a discrete time
Markov chain. The authors obtain results close to the simulations.



\subsection{Contributions and outlines}
In Section~\ref{s:norme}, we summarize the main characteristics of the
IEEE 802.11 standard when used in ad hoc networks with 802.11b devices.
We then present in Section~\ref{s:presentation} our chain of senders
scenario. Numerical values are given assuming a Lucent Orinoco 802.11b
wireless device. Comments of the three pairs fairness problem introduced
in \cite{Dhoutaut02} are also given.

In Section~\ref{s:extension}, we show by simulation using Network
Simulator, that interesting phenomena appear when varying the number $n$
of pairs: i) chance to gain access to the medium for the $i$th
sender-receiver pair depends on the parity of $i$, ii) the fairness
increases with $n$ especially for central pairs and iii) the system has
an asymptotic behavior when $n$ increases.

In Section~\ref{s:modeling}, we introduce a new modeling of such a
phenomenon. Although it is quite simple, it allows to match results of
simulations both for small and large values of $n$, depending on a
$\alpha$ coefficient. This coefficient corresponds to the probability of
emission when the neighbor senders are waiting. For small values of $n$,
we give close expressions (depending on $\alpha$) for the probability of
emission of a given pair.

In Section~\ref{s:analysis}, we prove that a stationary state exists for
each pair for any length of the chain. Moreover, this stationary state
converge to an asymptotic stationary state when $n$ increases. This
confirms the simulations.
We also show that some values of $\alpha$ allows to maximize the
fairness, expressed as entropy \cite{Jaynes57}.

In Section~\ref{s:discussion}, we comment these results, and we show
that when $n$ is large, the fairness is almost optimal near the center
of the chain. We also show that the simulation results tend to this
ideal case when $n$ increases.
Finally, we sketch the relationship between $\alpha$ and the IEEE 802.11
protocol, and we explain how to optimize the fairness by means of packet
size tuning relying on $n$ and $\alpha$.
As an application of our analytical study, we maximize the fairness in
the three pairs scenario. 

Concluding remarks end the paper.



\section{IEEE 802.11 standard in ad hoc mode}
\label{s:norme}

The IEEE 802.11 standard implements several types of wireless
communications \cite{IEEE802.11}. We focus on the most widely used for
ad hoc networking with 802.11b compliant devices in order to explain the
numerical values of this paper. We first begin by the physical layer and
then we summarize the medium access layer. Note that the numerical
values depend on the physical layer we describe, but this is not the
case for the fairness issues we point out, which appears also in
protocols based on other physical layer (such 802.11a or 802.11g for
instance).

\subsection{Physical layer}
In the 802.11 standard, the physical layer (PHY) is divided into two
sublayers: the \emph{Physical Medium Dependent} (PMD) covered by the
\emph{Physical Layer Convergence Sublayer}.

\subsubsection{PMD sublayer}
Besides the infra-red communications, the 802.11 PMD has been declined into two physical
layers for radio-communications, based on spread spectrum: FHSS and DSSS. The spread
spectrum techniques uses a wider bandwidth than needed for sending a message, leading to low
power density and redundancy: less energy is diffused on a given frequency causing less
interferences with the environment, and a given information is present in several
frequencies ensuring better noise robustness.
Others physical layers have been introduced in some addenda: HR-DSSS in
 \cite{IEEE802.11b} and OFDM in \cite{IEEE802.11a}. With the
 \emph{Channel Agility} option, a PMD can switch from one modulation to
 another.  However, ad hoc networks based on the IEEE 802.11b standard
 mainly rely on the DSSS and HR-DSSS PMD layers, operating in the
 2.4-2.485\,GHz frequency range included into the \emph{Industrial,
 Scientific and Medical} (ISM) frequencies. We know summarize these
 modulations.

For the \emph{Direct Sequence Spread Spectrum} (DSSS), the 2.4\,GHz ISM
range is divided into 14 channels of 22\,MHz each, with partial
overlapping. A single frequency is used for transmission. However a
chipping technique adds redundancy to increase the robustness: each bit
of data is coded by a sequence of eleven chips using a Barker code. The
modulation technique is the \emph{Differential Binary (resp.
Quadrature) Phase Shift Keying} (DBPSK, resp. DQPSK) which offers a
sending rate of 1\,Mbits/s for the DBPSK and 2\,Mbits/s for the DQPSK. In these
techniques, a phase rotation is performed depending on the symbol to
send (either one bit for DBPSK or two for DQPSK). These modulation
techniques admit a better minimum signal to noise ratio of about 12\,dB
than FHSS. However the transmission is more sensitive to multi-paths, and
to Bluetooth emissions (which uses the same bandwidth range).

The \emph{High Rate DSSS} (HR-DSSS) uses a more complex modulation
technique called \emph{Complementary Code Keying} (CCK). A sending rate
of 5.5\,Mbits/s (resp. 11\,Mbits/s) is reached with four (resp. eight)
symbols per chips. The different sending rates are chosen dynamically on
the basis of transmission conditions, for instance the signal to noise
ratio (this is not normalized). In practice, in outdoor environment, the
11Mbits/s is admissible until about 200\,m, the 5.5\,Mbits/s until about
300\,m, the 2\,Mbits/s until 400\,m and the 1\,Mbits/s until
500\,m. This of course depends on the devices (power), antenna (gain)
and environment (outdoor/indoor, obstacles, noise...).

\subsubsection{PLCP sublayer}
This sublayer makes a link between the different PMD layers and the MAC
layer (which should not depend on the physical layer, either current or
future). It prepares the MAC formated packets for the relevant PMD
layer. A header and a preamble are inserted before any sent data in
order to synchronize the sender and the receiver, to choose the
modulation technique, and so on.
As explained above, several data rates are available in the IEEE 802.11b
standard based on DSSS modulations: 1\,Mbits/s, 2\,Mbits/s, 5.5\,Mbits/s
and 11\,Mbits/s. While the norm admits the optional \emph{short
preamble and header} option (120 bits partially sent at 2\,Mbits/s
requiring 96\,$\mu$s), both preamble and header are generally sent at the
low sending rate (1Mbits/s using the DBPSK modulation) in order to be
understood by every stations (\emph{long preamble and header} default
option).

The (long) PLCP preamble is composed of 128 bits used for sender and
receiver auto-synchronization (SYNC field) and 16 bits for the
\emph{Start Field Delimiter} (SFD), that indicates the beginning of the
frame. This corresponds to 144~$\mu$s.
The (long) PLCP header is composed of the SIGNAL field (8 bits) to
indicate the modulation technique which is used (either DBPSK or DQPSK),
the SERVICES field (8 bits, currently unused), the LENGTH field (16
bits) to indicate the number of microseconds required for transmitting
the data of the MAC layer, and the CRC field (16 bits) used for the
cyclic redundancy code checking. This corresponds to 48~$\mu$s.
PLCP preamble and header lead to a total of 192~$\mu$s at the beginning of any sending.

The PLCD sublayer also implements the Carrier Sense/Clear Channel
Assessment (CS/CCA) procedure, which gives informations on the medium
(either idle or busy). It is used to detect the beginning of a network
signal which can be received (CS), and to determine whether the channel
is clear prior to transmit a packet (CCA). The duration of this
procedure depends on the modulation technique: 27\,$\mu$s for FHSS, less
than 15\,$\mu$s for DSSS and HR-DSSS. It impacts the value of the
\emph{aSlotTime} constant used by the MAC layer. By adding other
PHY-dependent delays, we found a slot time of 50\,$\mu$s for the FHSS
and 20\,$\mu$s for the DSSS and HR-DSSS modulations.

\subsection{Medium Access Control layer}
The purpose of the MAC layer is to control the access to the shared
medium by the neighborhood nodes. Two methods have been defined: the
\emph{Distributed Coordination Function} (DCF) and the \emph{Point
Coordination Function} (PCF). The fundamental access method is the DCF;
the PCF is optional. We focus on the DCF method which is the only used
in practice (PCF is rarely implemented). We first describe frames to
explain durations used in the rest of the paper.

\subsubsection{Frames}
A MAC frame is composed of a \emph{MAC header} (10 to 30 bytes, depending on the
kind of frame), a body (0 to 2312 bytes) and a \emph{Frame Check Sequence} (FCS,
4 bytes). The MAC header contains at least a \emph{Frame Control} field (2
bytes), a \emph{Duration} field (2 bytes) and a MAC address (6 bytes) leading to
a minimum frame size of 14 bytes with the FCS field and an empty body.
The header of a frame sent from one mobile to another one in an ad hoc network
is 24 bytes width.

Any frame is acknowledged by the receiver (unicast), implementing a
positive acknowledgment. If the acknowledgment has not been received
before a delay ACK\_TIMEOUT, the frame is sent again. An acknowledgment
is a 14 bytes length MAC frame (needing 304\,$\mu$s at 1\,Mbits/s when
adding the PLCP header and preamble).

\subsubsection{Delays}
The DCF implements a \emph{Carrier Sense Multiple Access} protocol with \emph{Collision Avoidance}
(CSMA/CA).  It is designed to reduce the collision probability by inserting some delays between
contiguous frames (\emph{interframe spaces}, IFS). The duration of the delay depends on the
situation.
Any transmission should begin by a \emph{DCF IFS} (DIFS) delay.
The acknowledgment is sent by the receiver after a \emph{Short IFS} (SIFS). The SIFS is smaller than
the DIFS to give priority to the acknowledgement to other transmissions.

If a station $S_2$ receives a frame but is not able to understand it
(erroneous frame), it waits during an \emph{Extended IFS} (EIFS) instead
of a DIFS before sending. This could be a frame sent by $S_1$ to $R_1$,
and these stations are too far from $S_2$ to allow a good reception by
this station (preamble and header are sent using the DBPSK modulation at
1\,Mbits/s, and can be understood while the rest of the frame sent at
higher rate with a different modulation could not be understood). The
EIFS delay allows to $R_1$ to acknowledge the frame sent by $S_1$. This
prevents some cases when $S_2$ does not hear the acknowledgment sent by
$R_1$, and begins a transmission that could prevent the acknowledgment
reception on $S_1$. The station $S_2$ will switch from EIFS to DIFS
delays after receiving a correct frame.

As for the aSlotTime constant, the duration of the SIFS delay depends on the PHY
layer. It is equal to 10\,$\mu$s for DSSS and HR-DSSS.
%
The DIFS delay is equal to a SIFS delay plus two aSlotTime, leading to
50\,$\mu$s for DSSS and HR-DSSS.
%
The EIFS delay is equal to a SIFS delay plus the duration of an
acknowledgment (sent at the lowest sending rate of 1\,Mbits/s) plus the
duration of a preamble and a header of the PLCP sublayer plus a DIFS
delay, leading to 364\,$\mu$s for DSSS and HR-DSSS.

\subsubsection{RTS/CTS}
Both physical and virtual mechanisms are available to sense the
carrier. As already seen, the PLCP sublayer provides a CS/CCA function
which is used by the MAC layer to probe the channel.  Moreover, each
station maintains a \emph{Network Allocation Vector} (NAV) in order to
foresee the channel liberation. The NAV is updated using the duration
field included in the received frames. A station cannot attempt to
transmit if its NAV indicates that the medium is busy. However a station
$S_2$ which is not in the neighborhood of the sender $S_1$ but is in the
neighborhood of the receiver $R_1$ could begin to send data during the
current transmission from $S_1$ to $R_1$, leading to a congestion on
$R_1$. To avoid this problem (\emph{hidden station}), the sender $S_1$
can first send a \emph{Request To Send} (RTS) message to $R_1$, which
will then reply by a \emph{Clear To Send} (CTS). The station $S_2$ will
receive the CTS message, and will then update its NAV, preventing it to
send data during the transmission $S_1 \rightarrow R_1$. The frames RTS
and CTS are followed by a SIFS delay.

A RTS frame has the same length than an ACK frame (14 bytes, 304\,$\mu$s
at 1\,Mbits/s). A CTS frame is 20 bytes long (352\,$\mu$s at 1\,Mbits/s)
because the header contains an additional MAC addresses. These frames
are supposed to be shorter than the data frames, and then less subject
to collisions. Depending on the configuration, this mechanism is i)
never used, ii) always used or iii) used when the frame length is larger
than a threshold.






\subsubsection{Backoff}
Despite the inter-frames delays and the carrier sense before any transmission,
several stations could decide to send simultaneously as soon as the medium is
clear. To minimize such a situation, any station waits for a random delay called
\emph{backoff time} before beginning a transmission.

After the DIFS or EIFS delay has expired, and if no current backoff time
remains, the station generates a random number $x$ between $0$ and the value of a
\emph{Contention Window} (CW). The backoff time is then equal to $x \times$aSlotTime.
Each time the channel is idle during aSlotTime microseconds, the backoff time is
decreased of aSlotTime microseconds. The backoff time does not decrease if the
medium is busy. The transmission can begin if the channel is idle and both the
delay (either DIFS or EIFS) and the backoff time has been expired.








The value of the contention window belongs to the interval CWmin and CWmax,
where CWmin depends on the physical layer (31 for DSSS and HR-DSSS) and CWmax
equals to 1023.
At the beginning, CW is equal to CWmin. Every time an attempt to transmit fails,
the contention window is doubled ($\text{CW} \leftarrow \text{CW} \times2 + 1$)
until it reaches CWmax.
The contention window is reset to CWmin after a successful transmission
(or after a fixed number of attempts). A successful transmission
includes an acknowledged frame as well as the receiving of a CTS frame
in response to a RTS frame.

\section{Fairness issues in a chain of senders}
\label{s:presentation}
The DCF mechanism described in the previous section ensures a fair
access to the shared medium when the competing nodes are able to hear
each of them. However in more complexe multi-hop ad hoc networks, some
cases of unfairness could be caused by asymmetry of the topology, or by
the use of the EIFS delay by some nodes while others use the DIFS
\cite{Nandagopal00,Chaudet05}. In this section, we present an unfair
case which appears in a chain of senders. This is a more general case
than the already known \emph{three pair problem} introduced in
\cite{Dhoutaut02}. We begin by some considerations on distances between
mobiles.

\subsection{Transmission ranges considerations}
In the 802.11 standard, the PHY layer reports the reception of a message only if
the \emph{Signal to Noise Ratio} (SNR) is larger than a fixed threshold
(SNR\_THRESHOLD).  A signal sent with a given transmission power will be
received with a smaller reception power because of signal attenuation, fading,
etc. This defines the \emph{transmission range} (Rtx) which is the maximal
distance to ensure a successful reception if there is no interference. The
transmission range mainly relies on radio propagation properties (attenuation),
and on the modulation technique used, that is on the environment and on the
sending rate.

As explained in the previous section, the PHY layer is also asked for
carrier sense detection (CS/CCA procedure). This mainly relies on the
antenna sensitivity. From a given distance called \emph{Carrier Sensing
Range} and denoted Rcs, the transmission of a far station is no more
detected.  Generally, the transmission range Rtx is smaller than the
carrier sensing range Rcs.
For instance, for a Lucent Orinoco wireless card, with a sending rate of
2\,Mbits/s, Rtx equals 400\,m while Rcs equals 670\,m \cite{Xu02}.


Suppose that a station $S_1$ sends a frame and a station $R_1$ tries to
receive it. For the reception to be feasible, we should have $d(S_1,R_1)
<$ Rtx where $d()$ denotes the Euclidean distance (here we admits an
outdoor environment). Now let us consider a third station $S_2$ further
from $R_1$ than $S_1$ that also sends some frames. On $R_1$, the
reception power of the signal sent by $S_2$ (denoted by $P_{r2}$) is
smaller than the one of $S_1$ (denoted by $P_{r1}$) and the signal of
$S_2$ is considered as noise. By comparing the ratio $P_{r1}/P_{r2}$ to
the SNR\_THRESHOLD, and by considering a signal attenuation in $1/d^4$
(corresponding to an outdoor environment modeled by the two-ray ground
propagation model outside the Fresnel zone), \cite{Xu02} determines an
\emph{interference range} Ri, which is equal to 1.78\,Rtx. This is the
maximal distance until which a station can disrupt a reception because
of concurrent sending.

These considerations lead to the following main cases (depicted on
Figure~\ref{f:range}), where the station $S_1$ sends some frames to $R_1$ while
another station $S_2$ could perturb this communication by its own emissions:

\begin{figure}[b!]
	\centering\includegraphics[width=0.8\columnwidth]{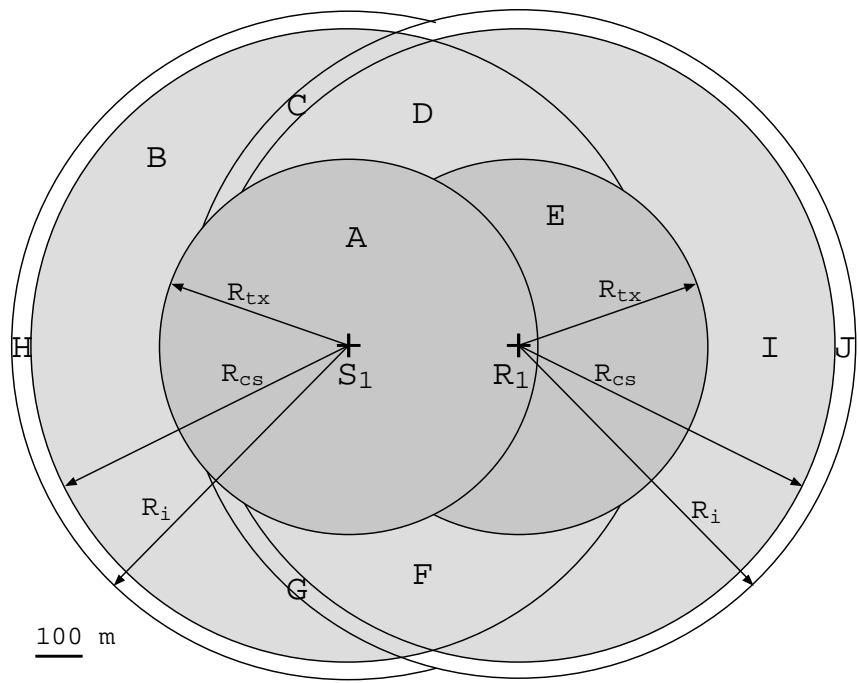}
	\caption{Communication ranges for a Lucent Orinoco 802.11b card in outdoor
	environment, with a sending rate of 2\,Mbits/s \cite{Xu02}.}\label{f:range}
\end{figure}

\begin{itemize}
\item If the station $S_2$ is in the area $A$, carrier sensing and backoff allow
to share the medium between $S_1$ and $S_2$.

\item If the station $S_2$ is in the area $E$, it is commonly called
\emph{hidden station} \cite{Kleinrock75}, and the RTS/CTS mechanism will
prevent the collision on $R_1$.

\item If $S_2$ is in the area $I \cup J$, the sending of $S_1$ and $S_2$
	will lead to some collisions on $R_1$ even if the RTS/CTS mechanism is
	used. Since $R_1$ will not acknowledge frames sent by $S_1$, $S_2$
	will increase its contention window.

\item If $S_2$ is in the area $B$, then it will receive the frames of
	$S_1$ without understanding them and will presume erroneous frames. As
	a consequence, it will wait an EIFS delay instead of a DIFS one,
	allowing $R_1$ to send the acknowledgment to $S_1$.
	
\item If $S_2$ is in the area $D \cup F$, then it will receive the
	frames of both $S_1$ and $R_1$ without understanding them and wait an EIFS
	delay.

\item If $S_2$ is in the area $C \cup G$, then it will receive the frames of
	$S_1$ without understanding them, and will use the EIFS delay. But it may
	also perturb the sending of some frames by $R_1$ (CTS and ACK), leading to
	a contention window increasing on $S_1$.

\item Finally, if $S_2$ is in $H$, then its sending will create some
	collisions on $S_1$ during the reception of the CTS and ACK frames
	sent by $R_1$, and $S_1$ will increase its contention window.
\end{itemize}

\subsection{Fairness in a chain of senders}

In this paper, we study the fairness in a chain of senders, where each
sender has one or several receivers which are not themselves senders (see
Figure~\ref{f:chain}): a sender continuously sends some data frames to
one of its neighbors, not necessarily always the same. As explained
previously, several kinds of interaction can appear between neighbor
senders and in some case their receivers. However many studies have
already be done on the increase of the contention window. In this paper, we
focus on the impact of the EIFS delay, which appears when a sender is in
the area $B \cup C \cup D \cup F \cup G$ (see Figure~\ref{f:range}) of
its neighbors, combined with the chain topology.

For the purpose of our study, we suppose that each sender is in the area
$B$. We noticed that very similar results are obtained when the sender
is in the area $D \cup F$, but the system stabilizes much
slowly. Moreover, as our simulations have been done with network
simulator \cite{ns} (see Section~\ref{s:extension}), the interferences
which may appear in the areas $C \cup G$ could not be taken into account.

This chain of senders scenario could rarely happen in a wireless LAN
network were the mobile nodes share an access point, because in such a
situation the stations are generally in the transmission range of either
the sender or the receiver (i.e.. $A \cup E$ in Figure~\ref{f:range}). But
it could appear more often in ad hoc network when the nodes are widely
spread in the space, and when they are moving.
More fundamentally, as we will see, this case study allows some interesting
conclusions on the IEEE 802.11 standard.

\subsection{The three pairs fairness problem}
In \cite{Dhoutaut02,Chaudet04b}, a specific scenario has been studied,
where strong inequity appears. It is based on asymmetry between some
pairs of communicating nodes, and on the use of the EIFS delay.
In this scenario, three pairs of communicating nodes are considered. In each
pair $i$ ($1 \leq i \leq 3$), the sender $S_i$ and the receiver $R_i$ are close
enough to establish a communication. Moreover, the sender $S_i$ has many data to
send to the receiver $R_i$ in the same pair so that it always tries to gain
access to the medium.
The three pairs are placed in such a way that the senders can detect an emission
in a neighbor pair without understanding the emission.

This is a particular case of our chains of transmitters scenario, as
depicted for instance in Figure~\ref{f:chain}. Here, there is a single
receiver per sender. These pairs of senders-receivers are not
necessarily arranged on a line, but a sender is in the carrier sense area of its
neighbors.

\begin{figure}[h!]
	\centering
	\epsfig{figure=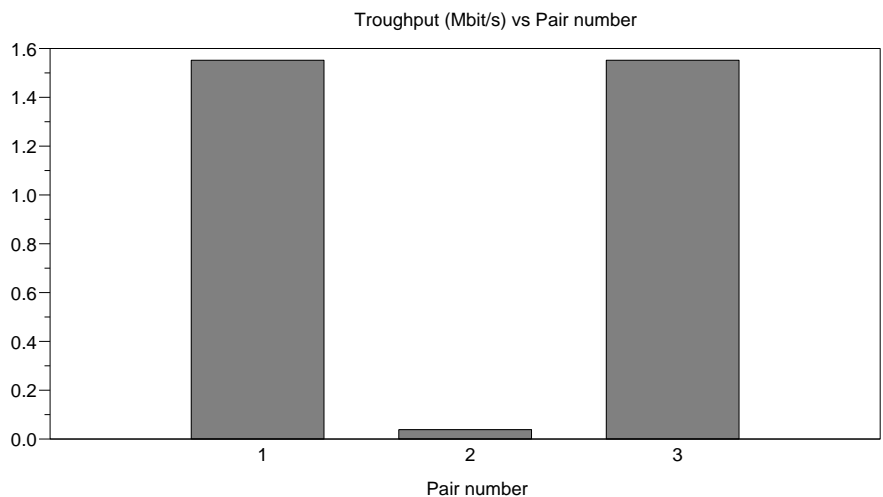,width=\columnwidth}
	\caption{Fairness problem with three pairs.\label{f:3pairs}}
 \end{figure}

Simulations have been done in \cite{Dhoutaut02} as well as real
experiments confirming the simulations.  Figure~\ref{f:3pairs} displays
simulations results of a chain of three pairs of senders-receivers, with
the parameters we will use in the following section. As already shown in
\cite{Dhoutaut02} (with different parameters), we notice a strong
inequity: the two external pairs can reach a throughput larger than
1.55\,Mbits/s, whereas the central pair has a throughput which does not
exceed 0.04\,Mbits/s. Note that in the same conditions, the throughput of
a single pair is equal to 1.59\,Mbits/s.

To explain these results, one can remark that the central pair has to
compete with two neighbors to access the channel, and then a smaller
throughput than the border pairs (which have only to compete with one
neighbor) is expected. Moreover, the EIFS mechanism applies as soon as a
neighbor is sending, and this happens more frequently for the central
pair.

\section{Simulation of a chain of senders}
\label{s:extension}

In the previous section, we introduced the chain of senders scenario,
which includes the three pairs fairness problem studied in
\cite{Dhoutaut02,Chaudet04b}. In such a scenario, the central pair has
many difficulties to gain access to the channel compared to its two
neighbors. But if those neighbors have more than one competitors, this
could help the central pairs. In the following we study by simulation
the impact of the number of pairs on the fairness in the chain of
senders. This scenario combines both the EIFS mechanism and the border
effect of the chain (some nodes have a single neighbor while some others
have two), which is expected to be less and less important when the
minimal distance to a border pair increases.

\subsection{Configuration and parameters}
\begin{figure}[h!]
	\centering
	\epsfig{figure=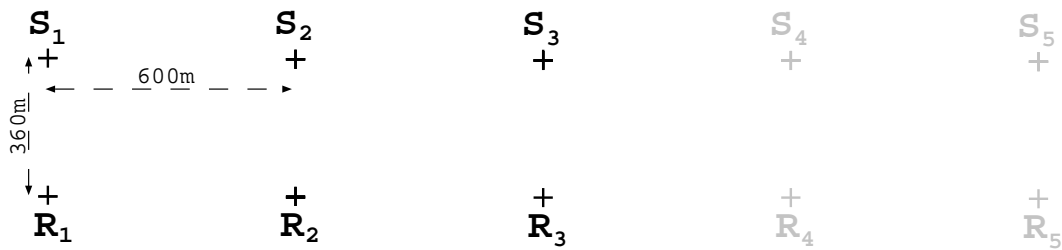,width=\columnwidth}
	\caption{Chain of sender-receiver used for the simulations.\label{f:chains2}}
\end{figure}

Our simulations have been done using Network Simulator v2.28 \cite{ns},
with parameters described previously and corresponding to a Lucent
Orinoco 802.11b device (see Section~\ref{s:norme} and
Figure~\ref{f:range}).  Without loss of generality, we assume a single
receiver per sender, leading to a chain of senders-receivers pairs.
These pairs are arranged as shown in Figure~\ref{f:chains2}. Similar
results should be obtained with a less regular pattern
(Figure~\ref{f:chain}), provided that the condition described in the
chain of senders scenario introduced in Section~\ref{s:presentation} are
fulfilled.

The data rate has been fixed to 2 Mbits/s, which corresponds to the
Figures~\ref{f:range} and \ref{f:chains2}. Each sender always tries to
send some UDP packets corresponding to a 1500 bytes MAC frame (see
Section~\ref{s:norme}), using the RTS/CTS mechanism. Note that we did
not notice a significant influence of RTS/CTS mechanism.
The propagation model is the \emph{two-ray ground}, corresponding to an
outdoor environment with a single reflection on the ground. Others
parameters are: transmission power (15~dBm), antenna height (0.9~m),
receiving threshold (-91dBm), carrier sense threshold (-100dBm)
\cite{Xu02}.  The next sections show some results when the number of
pairs is varying.


\subsection{Fairness in a chain of four pairs}
\begin{figure}[h!]
	\centering\includegraphics[width=\columnwidth]{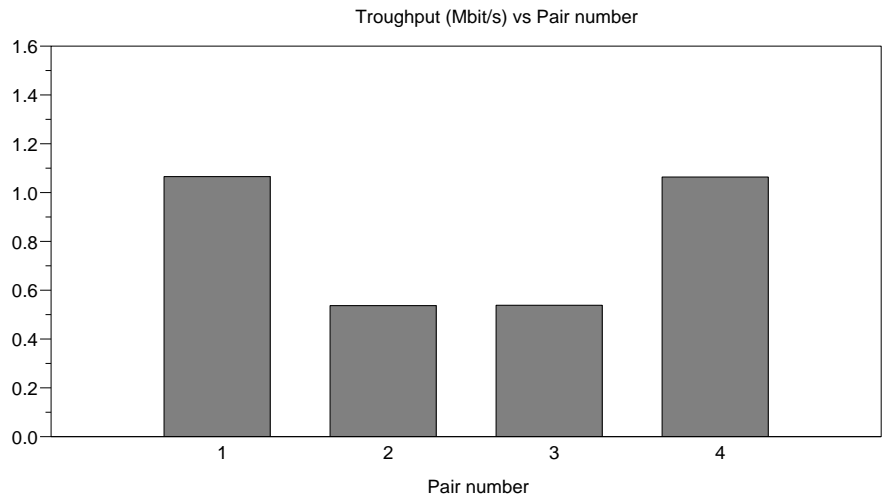}
	\caption{Fairness problem with four pairs.\label{f:4pairs}}
\end{figure}

Figure~\ref{f:4pairs} displays simulation results for a chain of four
pairs. We observe a different behavior than with three pairs
(Figure~\ref{f:3pairs}). The external pairs have a throughput around
1.06\,Mbits/s, whereas the two central ones reach only 0.53\,Mbits/s. As
previously said, this difference is explained by the number of
competitors: a single for the border pairs, and two for the central
ones.

Fairness is better than with three pairs because when the pair 1
acquires the channel, pair 2 is waiting and then pairs 3 and 4 have both
a single competitor. By comparison with the three pairs chain, when the
pair 1 acquires the channel, the other border pair always gains access
to the channel. Hence, with four pairs, the central pairs can have a more 
frequent access to the channel than the central pair in a chain of three
pairs.

Note however that when the pair 2 gains access to the channel, pairs 1
and 3 are waiting and then pair 4 acquires the channel without
difficulties. This explains the difference between central pairs and
border pairs.



\subsection{Fairness in a chain of five pairs}

\begin{figure}[h!]
	\centering\includegraphics[width=\columnwidth]{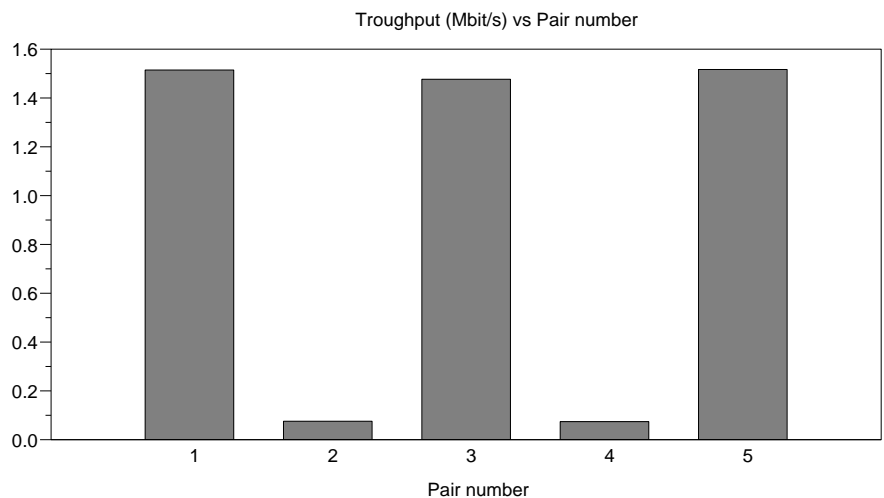}
	\caption{Fairness problem with five pairs.\label{f:5pairs}}
\end{figure}

Simulation results for five pairs are given in Figure~\ref{f:5pairs}.
As we can see, pairs 1, 3 and 5 have throughputs close to the maximum,
whereas pairs 1 and 2 have very low throughputs. 
Indeed, when the pair 1 gains access to the channel, the pair two is
waiting and the pairs 3, 4 and 5 have a similar behavior than a three
pair chain.

We observed a similar phenomenon with 7, 9 and 11 pairs.



\subsection{Fairness in a chain of six pairs}
\begin{figure}[h!]
	\centering\includegraphics[width=\columnwidth]{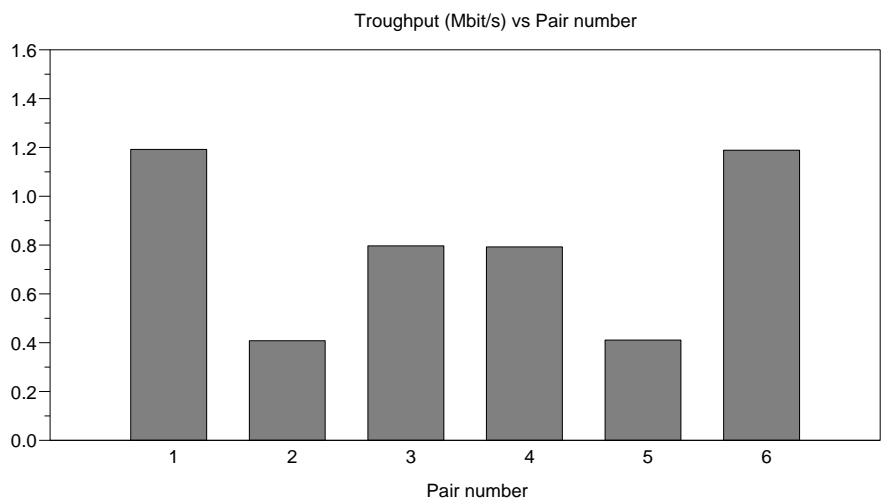}
	\caption{Fairness problem with six pairs.\label{f:6pairs}}
\end{figure}

Simulation results for six pairs are given in
Figure~\ref{f:6pairs}. They are not so far than results for four pairs,
except that pairs 2 and 5 have less bandwidth than central pairs 3 and
4, and that central pairs in the chain of four pairs. Here, even if the
border pair 6 acquires the channel, pair 2 could have more than one
competitor, which is not the case in a chain of four pairs. Note that
the pattern can also be seen as two neighbors chains of three pairs.

\begin{figure}[h!]
	\centering\includegraphics[width=\columnwidth]{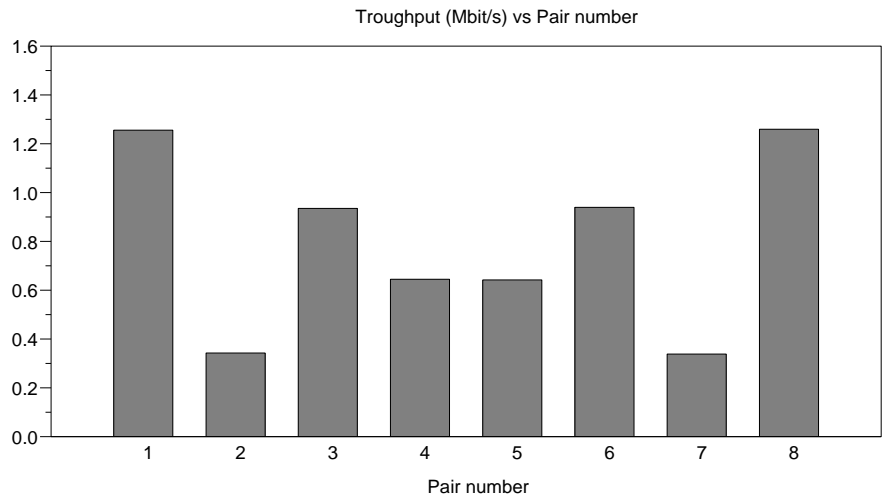}
	\caption{Fairness problem with eight pairs\label{f:8pairs}}
\end{figure}

We observed some similar behaviors for the chains with a larger even
number of pairs, as seen in Figure~\ref{f:8pairs} with eight pairs.

\subsection{Fairness in a chain of one hundred pairs}
As explained below, the fairness pattern in a chain of $n$ pairs depends
on the parity of $n$, which is an interesting phenomenon.  When $n$ is
odd, the fairness is bad (Figures~\ref{f:3pairs} and
\ref{f:5pairs}). When $n$ is even, some more complex patterns appear
with better fairness (Figures~\ref{f:6pairs} and \ref{f:8pairs}).

However we also observed some evolutions of these patterns when $n$
increases. We then simulated a very large chain, in order to have an
idea of the asymptotic behavior.

\begin{figure}[H]
	\centering\includegraphics[width=\columnwidth]{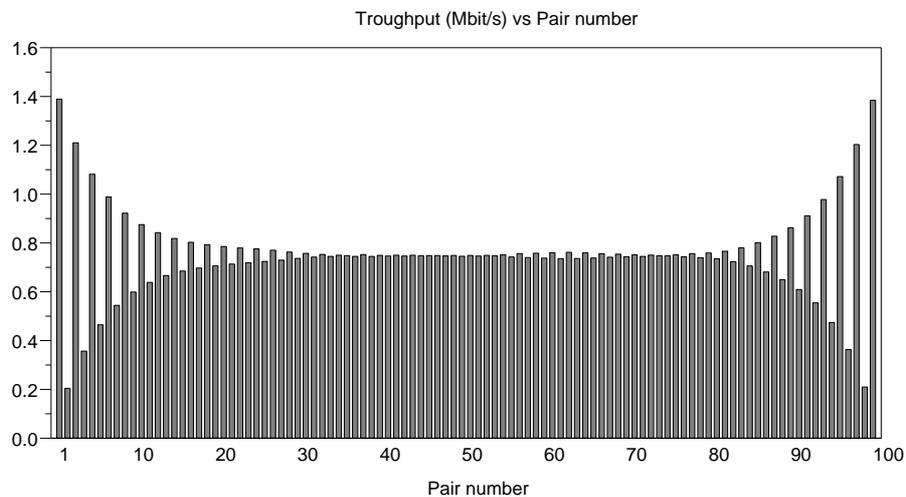}
	\caption{Fairness problem with one hundred pairs.\label{f:100pairs}}
\end{figure}

Figure~\ref{f:100pairs} displays the simulation results for a chain of
one hundred of pairs. We observed that the same result is obtained with
a chain of 101 pairs, which confirms that the influence of the parity of
$n$ tends to decrease when $n$ increases.
Moreover, for a chain of 101 pairs, one can see that the closer is an
even pair from the middle, the larger is its throughput. This is
explained by the fact that the influence of the border pairs is less
important. As a consequence, the closer is an even pair from a border,
the smaller is its throughput.

In this chain, the throughput of external pairs (1.39\,Mbits/s) is very
close to the maximum (1.59\,Mbits/s), measured in a single pair in the
same conditions. In the central flat area, the throughput of the pairs
is close to 0.75\,Mbits/s (about half of the throughput of the external
pairs). As a consequence of the existence of this flat area, the
insertion of a new pair has less influence on the throughput of other
pairs when $n$ is large, and when the new pair is inserted near the
middle of the chain.


\section{Mathematical modeling}
\label{s:modeling}

\newenvironment{myequation}
{\begingroup%
\equation}
{\endequation\endgroup}
\newenvironment{mydisplaymath}
{\begingroup%
\displaymath}
{\enddisplaymath\endgroup}

In the previous section, we have shown that a chain of senders presents
some interesting phenomena, depending on the number $n$ of pairs in the
chain, and on the parity of $n$. The three pairs fairness problem
introduced in \cite{Dhoutaut02} appear as a sub-case of the chain of
senders scenario presented in Section~\ref{s:presentation}.

In this section, in order to study this phenomena and to improve the
fairness, we propose a simple modeling of such a phenomenon, before
comparing the model with the simulations.

\subsection{Modeling with a non-linear system of equations}
In \cite{Chaudet04b}, a mathematical modeling has been proposed for the
three pairs configuration, by means of discrete time Markov chains.
Such a modeling gives numerical results close to the simulations
obtained with the ns-2 network simulator, and not so far from real
experiments of \cite{Dhoutaut02}. Moreover, it allows to study the
influence of some parameters variations on the fairness.
However, it is not easily generalizable when the number of pairs
increases. Indeed, a state of the Markov chain needs to capture the
relative remaining backoff delays of the pairs, which leads to many
states. Moreover, transitions are more complex when the number of pairs
(and then interactions) increases.

We propose a new modeling, based on a non-linear systems of $n$
equations whose solution gives the probabilities of emission of each
pair. It allows an analytical study both for small and large values of
the number $n$ of pairs.

Let us consider a chain of $n$ pairs numbered from $1$ to $n$. For the
purpose of the modeling, we admit that there are two border pairs (pair
$0$ and $n+1$), which never send data.

We consider the random process $y_i(t)$ taking value $1$ if the
$i^\mathrm{th}$ pair is sending data at time $t$ and 0 if the pair is
idle. In fact for any $t$, the random variable $y_i(t)$ follows a
Bernouilli's law. We now make a simple analysis of the communication
mechanism in order to obtain some relationships between the variables
$y_i(t)$, for $i=1\dots n$.

Some data can be sent in a given pair $i$ only if its neighbor pairs are
idle. Thus we have the implication
\begin{equation}
y_i(t)=1 \Longrightarrow y_{i-1}(t)=y_{i+1}(t)=0.
\label{eq:impl}
\end{equation}
But before emitting, the sender first waits after delays and CTS frames,
so the converse of (\ref{eq:impl}) is not true. To take this into
account, we introduce a new random process $z_i(t)$ such that
$$
P\left(z_i(t)=1\left\vert y_{i-1}(t)=y_{i+1}(t)=0\right.\right)=\alpha,
$$
where $0<\alpha<1$ and we consider that data can be sent in pair $i$ at time $t$ if neighbor pairs are idle and $z_i(t)=1$. Thus we can write the algebraic relationship
\begin{equation}
y_i(t)=z_i(t)\left(1-y_{i-1}(t)\right)\left(1-y_{i+1}(t)\right),~i=1\dots n.
\label{eq:relation}
\end{equation}
Since we want to describe some average behavior, we consider the rate of emission as the limit when $T\rightarrow \infty$ of the
time elapsed in the emitting state between $t=0$ and $t=T$ divided by $T$
$$
x_i=\lim_{T\rightarrow \infty}\frac{1}{T}\int_0^Ty_i(t)\,dt.
$$ In virtue of the Limit Central Theorem we have $x_i=E[y_i(t)]$, where
$E[.]$ denotes the mathematical expectation, and we have of course
$$
E[y_i(t)]=P(i\text{ is emitting at time }t),
$$ 
since $y_i(t)$ follows a Bernouilli's law. Hence, we can take the mathematical expectation on both sides of (\ref{eq:relation}), and we obtain, by neglecting the correlation between pairs $i-1$ and $i+1$
\begin{equation}
x_i=\alpha (1-x_{i-1})(1-x_{i+1}),~i=1\dots n.
\label{eq:proba}
\end{equation}

\subsection{Analytical results}
The modeling introduced above allows to obtain, by substitution of
unknowns and by using symmetry relationships, a closed form of
probabilities of emission, at least for small values of $n$. For
instance, for $n=3$, we have:
\begin{mydisplaymath}
x_1 = \frac{2\alpha^2-1+\sqrt{(1-2\alpha^2)^2-4\alpha^3(\alpha-1)}}{2\alpha^2}
\end{mydisplaymath}

For $n=4$, we have:
\begin{mydisplaymath}
x_1 = \frac{1+\alpha - \sqrt{(1-\alpha)(1+3\alpha)}}{2\alpha}
\end{mydisplaymath}

Similar expressions can be found for other pairs, but for $n>8$, there is no analytical formula because using the substitution technique leads to a polynomial with degree greater or equal than $5$, and the solution of (\ref{eq:proba}) has to be computed with numerical techniques.

\subsection{Validation with ns-2 results}\label{s:modeling-validation}
In order to compare these results with those given by the ns-2 network
simulator, we normalize both results by the value of the first external
pair. Indeed, during a period of $t$ seconds, the $i^\mathrm{th}$ pair
can send data during $t_i = x_i \times T$ seconds. Let $r_i$ be the
sending rate of the $i^\mathrm{th}$ pair determined by ns-2, in
bits/seconds. We have $r_i \times T = r_{max} \times t_i$ where
$r_{max}$ represents the maximal sending rate depending on the
configuration and $t_i$ the total time during which the $i^\mathrm{th}$
pair has sent data. Thus $r_i / t_i$ is a constant equal to $r_{max}
/T$, and we have $r_i/t_i = r_1/t_1$ and then $r_i/r_1 = t_i/t_1 = x_i /
x_1$. We then compare the throughputs of each pair divided by the
throughput of the first one ($r_i/r_1$) with the probability of emission
of each pair divided by the probability of emission of the first one
($x_i/x_1$).

\begin{figure}[H]
	\centering\includegraphics[width=\columnwidth]{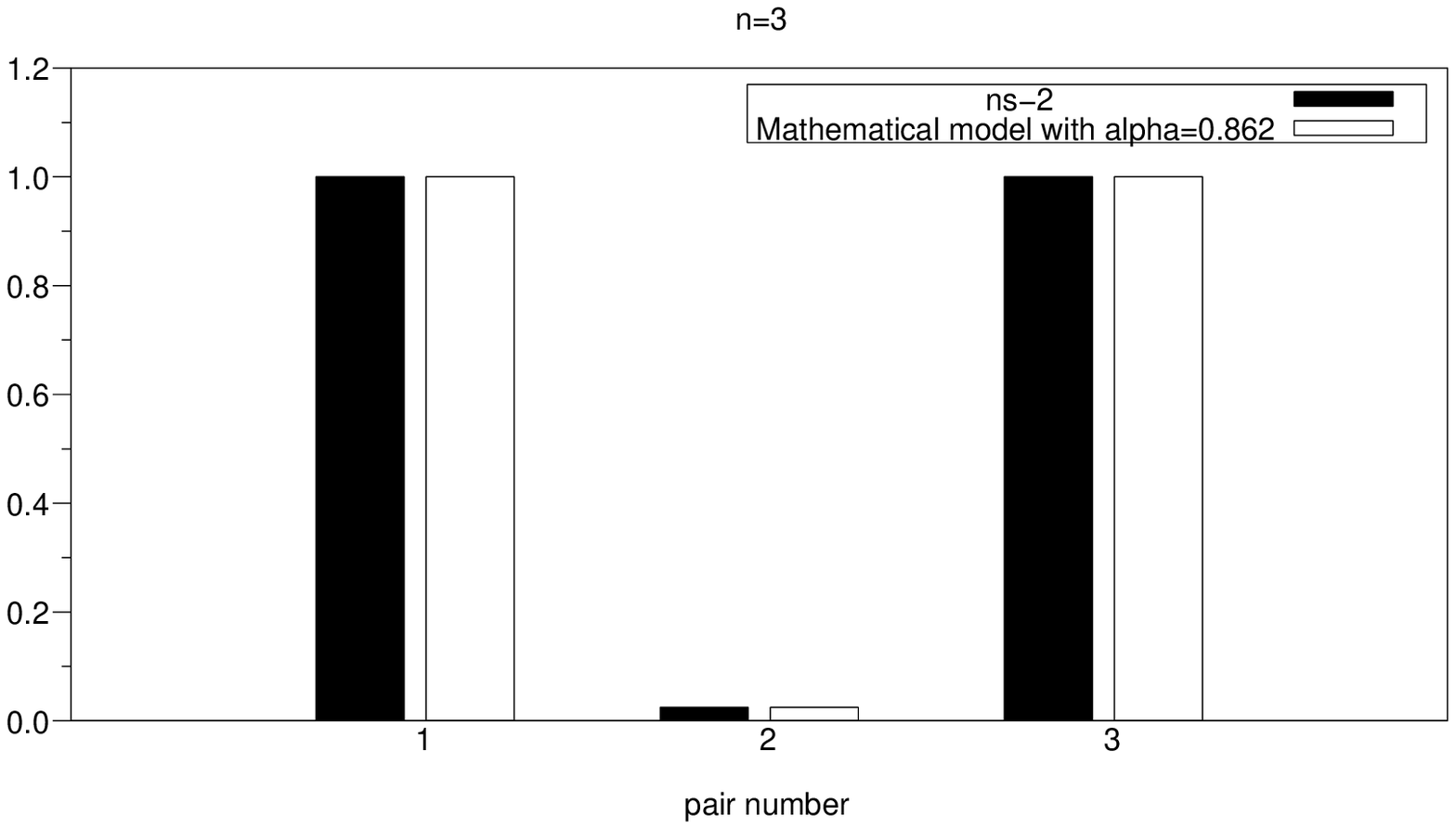}
	\centering\includegraphics[width=\columnwidth]{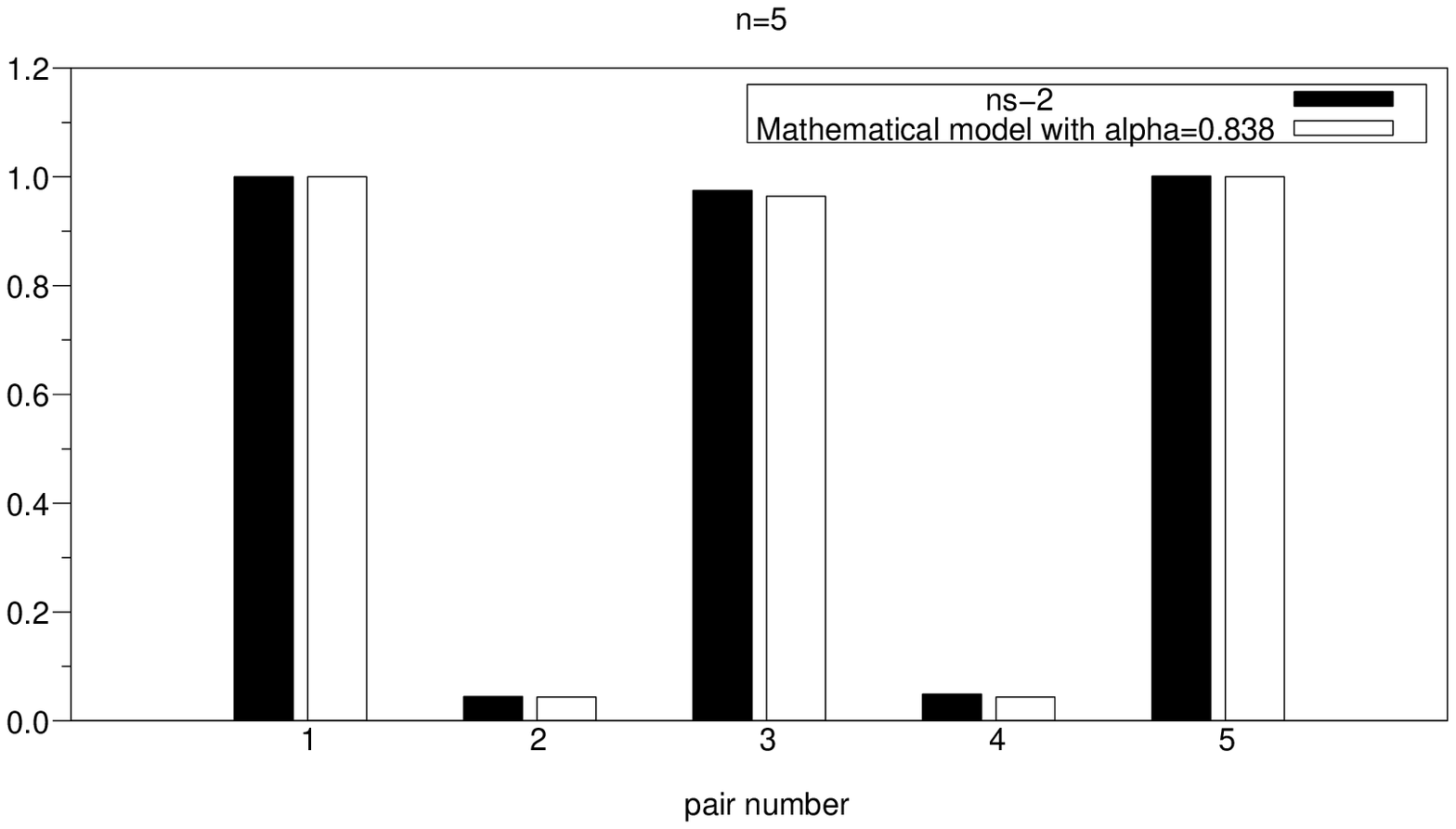}
	\centering\includegraphics[width=\columnwidth]{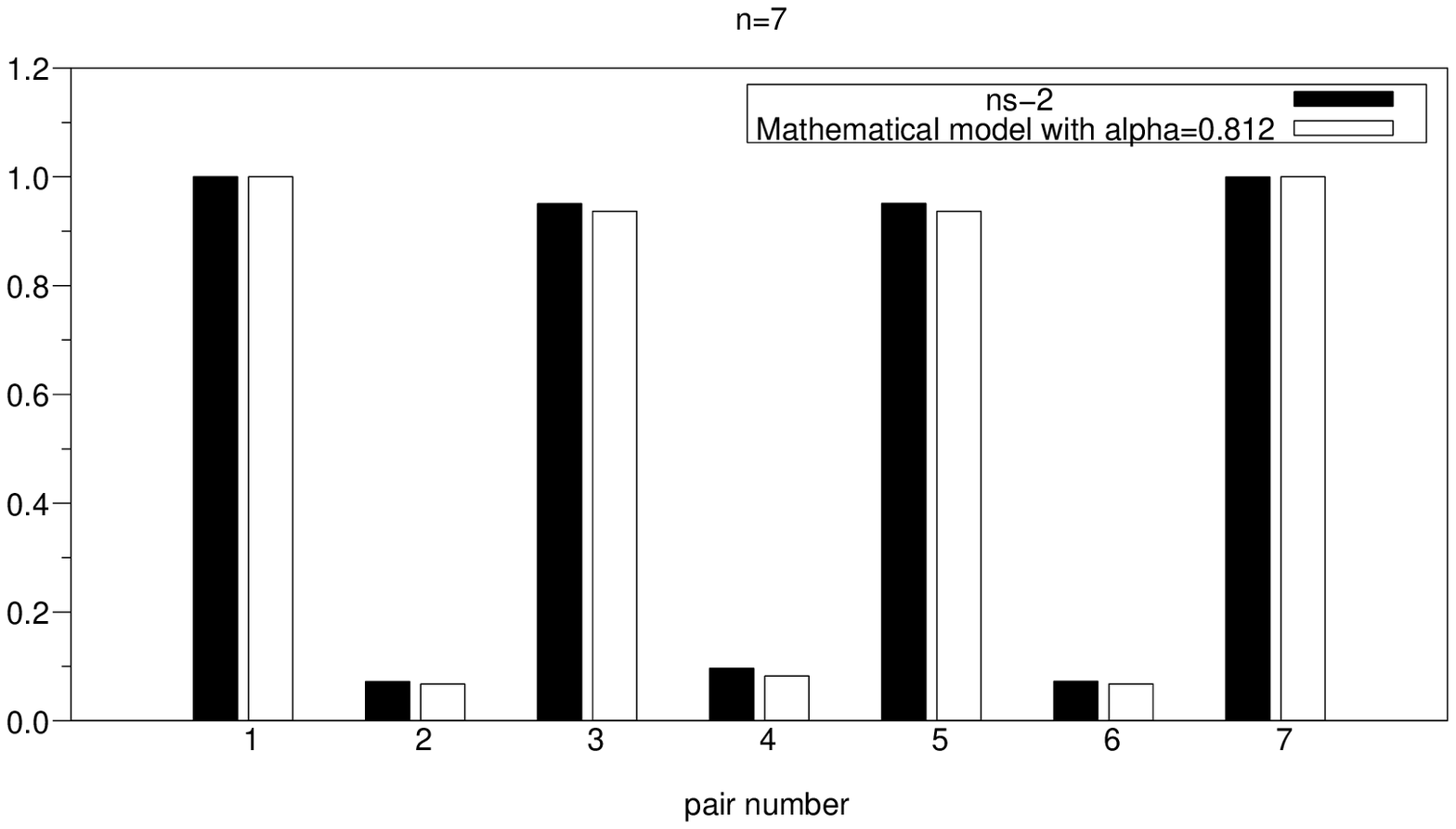}
	\caption{Comparison of $ns-2$ results and mathematical modeling for
	$n=3$, $5$ and $7$.\label{f:cmp}}
\end{figure}

We have done a least squares fitting with respect to $\alpha$ to
approximate the ns-2 results. For instance, for $n=3$, $n=5$ and $n=7$,
we obtain values of $\alpha$ respectively equal to $0.862$ and $0.838$
and $0.812$.  These values of $\alpha$ lead to numerical results very
close to those obtained with ns-2 network simulator, as seen in
Figure~\ref{f:cmp} (a discussion of these values is given in
Section~\ref{s:discussion}). The slightly differences are insignificant
compared to the unavoidable approximations of the network simulator.
Nevertheless, this first observation is only a rough validation of our
modeling, and a precise analysis of the model itself is necessary.

\section{Analysis of the model}\label{s:analysis}

Our simple modeling of the chain of senders scenario fits very well with
the simulations results for some given values of $\alpha$ (that we
discuss in Section~\ref{s:discussion}). In this section, we use this
modeling to determine the asymptotic behavior of the chain, as well as
to establish the relationship between $\alpha$ and the fairness.

\subsection{Proving the existence of a solution}

Let us consider the $n$ values $x^{k}_1 \ldots x^{k}_n$ as the
components of the vector $x^{(k)}\in \mathbb{R}^n$, and the iterative
process by means of a function $F_\alpha$ defined on vectors:
\begin{equation}
x^{(k+1)} = F_\alpha(x^{(k)}).
\label{successive-approximation}
\end{equation}

We have:
\begin{equation}
F_\alpha(x) = \alpha \left( \begin{array}{c}
1 - x_2\\
(1 - x_1)(1 - x_3)\\
\vdots \\
(1 - x_{n-2})(1-x_n)\\
(1- x_{n-1})                            \end{array} \right).
\label{eq:Falpha}
\end{equation}

The algorithm (\ref{successive-approximation}) is nothing but the so-called {\em successive approximation method} to 
determine iteratively a solution of the equation $x=F_\alpha(x)$. The convergence toward a unique solution $\hat x\in E$ is guaranteed provided the application $F_\alpha:E\rightarrow E$ is a contraction in some domain $E\subset \mathbb{R}^n$ (this is the well-known "contraction mapping theorem", see \cite{Rudin}). To show that $F_\alpha$ is a contraction we can use, since $F_\alpha$ is differentiable, 
the derivative $F_\alpha'$ given by the matrix
\begin{mydisplaymath}
F_\alpha'(x)=\alpha\left( \begin{array}{ccccc}
	0              & -1 & 0                & 0       & 0 \\
  x_3-1 & 0       & x_1 - 1 & 0       & 0 \\
                 & \ddots  & \ddots           & \ddots  & \\
                 &         & 1-x_n   &  0      & 1-x_{n-2}\\
  0              & 0       & 0                & -1 & 0
							 \end{array}\right).
\end{mydisplaymath}
If we take the supremum norm, i.e. $\Vert x\Vert=\max_{1\leq i\leq n}\vert x_i\vert$, we can show that $\Vert F_\alpha'(x)\Vert<1$, provided that $\vert x_k-1\vert < \frac{1}{2\alpha},\;1\leq k\leq n$, i.e. $F_\alpha$ is a contraction on the subspace $E$ defined by 
$E=\{x\in\mathbb{R}^n,\;\Vert x-\mathbf{1}\Vert < \frac{1}{2\alpha}$, where $\mathbf{1}=(1,\dots,1)$.

A direct application of this result is that the algorithm (\ref{successive-approximation}) converges to the unique solution of $x=F_\alpha(x)$ e.g. by taking $x^{(0)}=(1,\dots,1)$.

\subsection{Asymptotic behavior}

\begin{figure}[h!]
	\centering\includegraphics[width=\columnwidth]{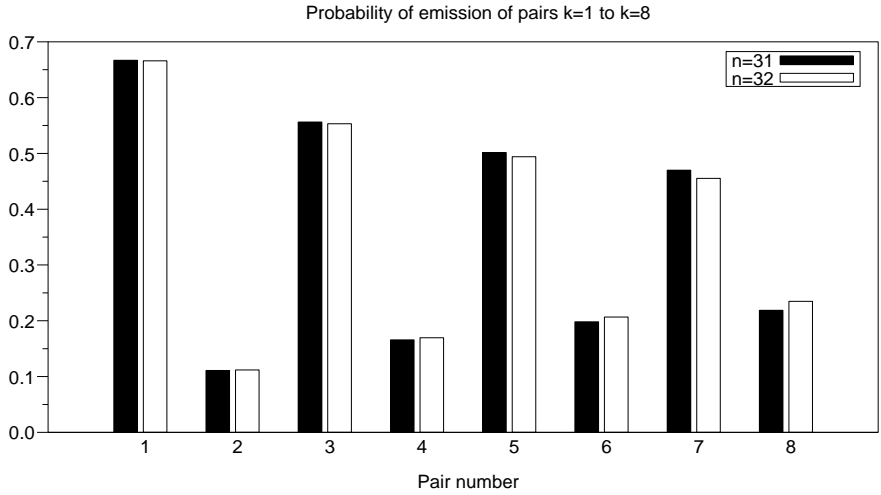}
	\caption{Simulation of probabilities of emission of pairs $k=1$ to $8$ for $n=31$ and $n=32$ ($\alpha=0.75$)\label{fig:rates15and16}}
\end{figure}

As for the simulations, we observe the convergence to an asymptotic
behavior. And the different behaviors between odd and even values of $n$
tend to disappear when $n$ increases.  Figure~\ref{fig:rates15and16}
shows the probability of emission of pairs $k=1$ to $8$ for $n=31$ and
$n=32$, for $\alpha=0.75$. For much greater values of $n$, the
difference between the rates of the first $n/2$ pairs for $n$ (even) and
$n+1$ pairs is negligible (typically less that $10^{-5}$ for
$n=100$). Thus, without loss of generality, we will continue our study
by considering only even values of $n$ in the simulations.

\subsection{Maximization of fairness with respect to $\alpha$}

Among other possible criteria (see \cite{Koksal00} and \cite{BM01}), one
way of maximizing the fairness between all pairs is to maximize the
entropy (see \cite{Jaynes57}) of the distribution of probability of
emission $\{x_i\}_{i=1\dots n}$, i.e. the function
$$ E(x)=-\sum_{k=1}^n x_i\log x_i.
$$
Hence, we consider the function $J(\alpha)=\frac{1}{n}E(x(\alpha))$ where $x(\alpha)$ is the unique solution of the equation $x=F_\alpha(x)$ and the factor $\frac{1}{n}$ is used to allow some comparisons of results between different values of $n$.

We search for the value $\hat \alpha$ such that
\begin{equation}
J(\hat \alpha)\geq J(\alpha),\;\forall \alpha\in[0,1].
\label{eq:opt}
\end{equation}

\begin{figure}[htb]
		\centering
    \includegraphics[,width=\columnwidth]{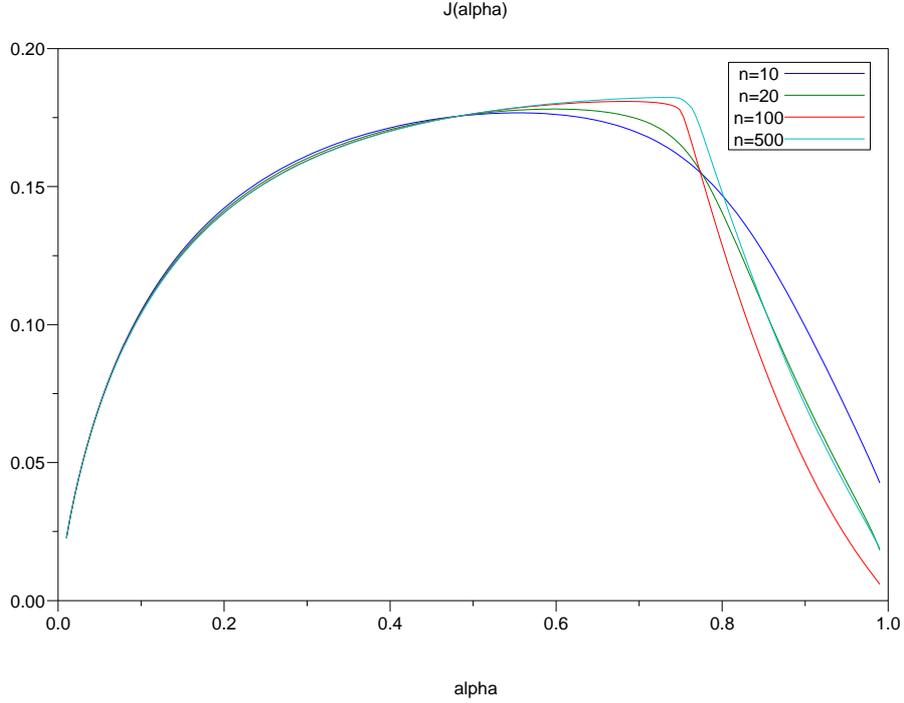}
	\caption{$J(\alpha)$ with respect to $\alpha$ for $n=10,20,100$ and $500$\label{f:Jalpha}}
\end{figure}

The Figure \ref{f:Jalpha} represents $J(\alpha)$ with respect to $\alpha$ for $n=10,20,100$ and $500$. For these values of $n$ we have respectively $\hat \alpha=0.5536,0.5977,0.6826,0.7309$.

The derivative of $J(\alpha)$ with respect to $\alpha$ is computed by using the classical adjoint state method, i.e. we consider the
Lagrangian
$$
L(\alpha,x,\lambda)=\frac{1}{n}E(x) +\lambda^\top (x-F_\alpha(x)),
$$ where $\lambda$ is a vector of $\mathbb{R}^n$ and $\top$ denotes the
transposition. The function $F_\alpha$ has been defined in Equation
(\ref{eq:Falpha}). We have, of course,
$J(\alpha)=L(\alpha,x(\alpha),\lambda)$ for any $\lambda$. We choose
$\lambda=\lambda(\alpha)$ such that
$$
\frac{\partial L}{\partial x}(\alpha,x(\alpha),\lambda(\alpha))=0,
$$
which leads to $\lambda(\alpha)=\frac{1}{n}[F_\alpha'(\alpha)-I]^{-1}\nabla E(x(\alpha))$, where $\nabla E$ is the gradient of $E(x)$ with respect to $x$.
We have finally
\begin{eqnarray}
J'(\alpha)&=&-\lambda(\alpha)^\top\left(\frac{\partial L}{\partial \alpha }F_\alpha(x(\alpha))\right),\\
&=&-\frac{1}{\alpha}\lambda(\alpha)^\top F_\alpha(x(\alpha)).
\end{eqnarray}

The computation of $x(\alpha)$ is done with a Newton type method, much
faster than the simple fixed point method suggested by Equation
(\ref{successive-approximation}), and the optimization is performed by
the Quasi Newton BFGS method available in Scilab (see \cite{Scilab99}).

\section{Discussion}\label{s:discussion}

In the previous section, the chain of senders scenario has been analyzed
on the basis of the modeling introduced in
Section~\ref{s:modeling}. Note that as far as the mathematical model is
concerned, the non-linear systems of equations (\ref{eq:proba}) is
obtained by assuming that the emission states of pairs $i$ and $i+1$ are
independent from a probabilistic point of view. While this assumption
(also assumed in \cite{Wang04}) may be questionable, it is relevant
because our modeling considers the stationary behavior of the chain.

In this section, we discuss the asymptotic values obtained in the
analysis before interpreting $\alpha$ in a practical point of view.

\subsection{Asymptotic flat area}

\begin{figure}[h!]
	\centering\includegraphics[width=\columnwidth]{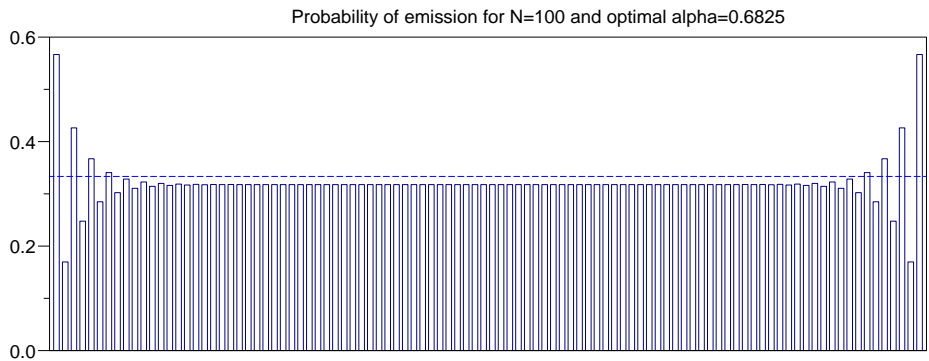}
	\caption{Probabilities of emission for $n=100$ and optimal $\alpha$. The dotted line is at
	probability $1/3$.\label{fig:opt100}}
\end{figure}

If we study the asymptotic behavior of results, we see that for large
values of $n$ and the optimal value $\alpha=\hat \alpha$, the optimal
probabilities of emission (see Figure \ref{fig:opt100}) exhibit a large
flat area with a value very close to $\frac{1}{3}$ (the $\frac{1}{3}$
value will be discussed below). This flat area ensures that the
insertion of a new pair will not disturb the rate for close neighbors.

Moreover, for $n=100,500,1000$ and $2000$ the value of the optimal
probability corresponding to this flat area is respectively equal to
$0.3177$, $0.3290$, $0.3313$ and $0.3325$.

To understand the convergence of this value to $1/3$, we must consider
the idealized situation where there is an infinite number of pairs, or
equivalently, the situation where the number of pairs is large enough to
allow to form a circle, where the last pair numbered $k=n$ has the pairs
$k=n-1$ and $k=1$ as neighbors. Hence, there is no border effect since
all pairs have two neighbor pairs.

So let us consider the $i^\mathrm{th}$ pair and its neighbors pairs
numbered $i-1$ an $i+1$, and a very simple model of channel acquirement:
each sender of each pair generates a realization of a random variable
$u_i$ (uniformly distributed in the interval $[a,b]$). We consider that
the $i^\mathrm{th}$ pair will acquire the channel if $u_i<u_{i+1}$ and
$u_i<u_{i-1}$. The probability of this event can be calculated as
follows:
\begin{eqnarray*}
P(u_i<u_{i+1},u_i<u_{i-1})&=&\int_a^b\int_a^{u_{i}}\int_a^{u_i}\,\frac{d_{u_{i+1}}
d_{u_{i-1}}d_{u_{i}}}{(b-a)^3},\\ &=&\frac{1}{(b-a)^3}\int_a^b (u_i-a)^2\, d_{u_{i}},\\&=&\frac{1}{3}.\end{eqnarray*}

Hence, the value $\frac{1}{3}$ can be understood as a limiting value
exhibiting the maximum fairness that can be obtained. This value of
$\frac{1}{3}$ is asymptotically obtained in our model, by maximizing the
entropy of the distribution of probabilities: this is a very interesting
behavior.

\subsection{Asymptotic optimal alpha}

\begin{figure}[h!]
	\centering\includegraphics[width=\columnwidth]{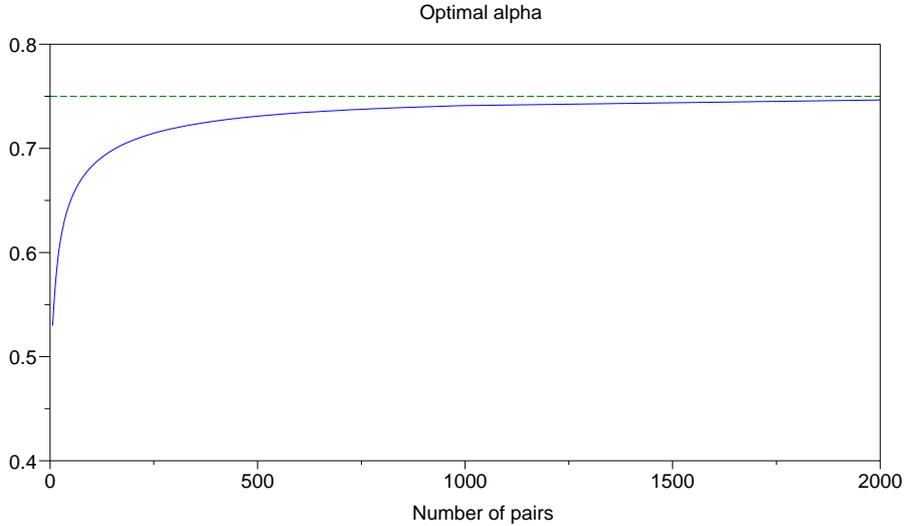}
	\caption{Optimal $\alpha$ with respect to $n$.\label{f:limiteAlpha}}
\end{figure}

Another interesting phenomenon is the apparent convergence of the
optimal value $\hat \alpha$ to $0.75$ when $n$ tends to the infinity, as
it can be seen on Figure \ref{f:limiteAlpha}.

This is not so surprising, as we will show it in the following
analysis. Consider the same idealized situation as before, where the
pairs are arranged to form a circle: the probabilities of emission
$\{x_k\}_{k=1\dots n}$ are necessarily invariant with respect to a shift
of indices, since all pairs will always have two neighbors. Hence we
have $x_k=x_1$, $\forall k$, and the system of $n$ equations
$x=F_\alpha(x)$ giving the probabilities is equivalent to the scalar
equation $x_1=\alpha(1-x_1)^2$. In this case the entropy is already
maximized since all values are equal. Then, if we are looking for the
value of $\alpha$ giving the maximum probability of emission in such a
configuration, i.e. $x_1=\frac{1}{3}$, we obtain
$\alpha=\frac{x_1}{(1-x_1)^2}=0.75$. This value is in fact completely
determined by the topology of the neighborhood.

\subsection{Asymptotic comparison of modeling and simulation}
\begin{figure}[h!]
	\centering\includegraphics[width=\columnwidth]{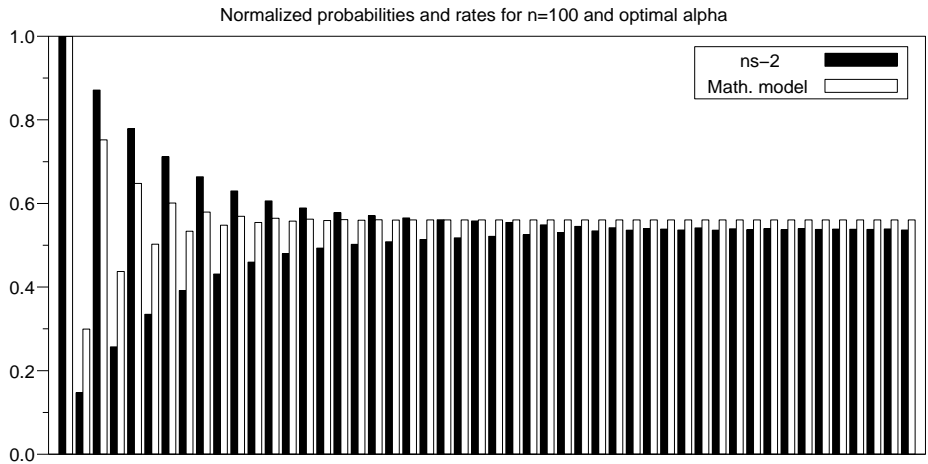}
	\caption{Probability of emission of the first $50$ pairs of 100
	obtained by ns-2 and mathematical model for optimal
	$\alpha=0.6825$.\label{f:nsVSmath}} 
\end{figure}

We have compared the normalized rates obtained via ns-2 and via the
mathematical model for $n=100$ pairs (the rates and probabilities are
normalized with respect to the pair exhibiting the maximum value, as
explained in Section~\ref{s:modeling-validation}). On
Figure~\ref{f:nsVSmath} we can see that the mathematical model with
$\alpha=0.6825$, corresponding to the maximum entropy, gives an
excellent approximation of ns-2 results.

Hence, it appears that the asymptotic behavior of the chain of $n$ IEEE
802.11 senders-receivers (as defined in Section~\ref{s:extension}) tends
to the maximum entropy when $n$ tends to the infinity. This is a
surprising result.

\subsection{Interpretation of the $\alpha$  coefficient}

We defined $\alpha$ as the probability of sending for a given pair when
its neighbors are not sending. Interpreting $\alpha$ implies to
determine whether a pair is sending or not when its neighbors are not
sending. This in fact depends on what is able to hear a neighbor sender,
and then on what area it is on Figure~\ref{f:range}. As for previous
simulations, we suppose that the neighbors senders are in the area $B$,
and that a sender can only hear transmission of a neighbor sender, and
not of a neighbor receiver. A neighbor pair is then considering as
sending only when the sender (and not the receiver) is sending, and
waiting in other cases.

Before any transmission, a sender has to wait for a delay, and in many
cases this is an EIFS delay instead of a DIFS one. During this delay,
chances are large that its neighbors are sending. This means that this
delay is not part of the time wasting by a pair while it could send
because its neighbors are not sending. To the contrary, neighbor senders
are not sending during the backoff delay.

Figure~\ref{f:transmission} summarizes a complete transmission of a $s$
bytes MAC frame between a sender $S_i$ and a receiver $R_i$ using
numerical values given in Section~\ref{s:norme} ($d$ denotes the sending
rate, and 0.5 represents the mathematical expectation of a random
variable on $[0,1]$).

\begin{figure}[h!]
\centerline{
\small
\begin{tabular}{|c|c|c|c|}
\hline
\multicolumn{2}{|c|}{sender $S_i$} & \multicolumn{2}{|c|}{receiver $R_i$} \\
\hline
DIFS or EIFS                   & 50 or 364 $\mu$s      &      & \\
aSlotTime $\times$ CW $\times 0.5$ & 310$\mu$s  &      & \\
RTS                            & 304 $\mu$s            &      & \\
                               &                 & SIFS & 10 $\mu$s\\
                               &                 & CTS  & 352 $\mu$s\\
SIFS                           & 10 $\mu$s             &      &     \\
header and preamble (PHY)      & 192 $\mu$s            &      & \\
$s$ data bytes (MAC)           & $8 \times s /d$ $\mu$s &      & \\
                               &                 & SIFS & 10 $\mu$s\\
                               &                 & ACK  & 304 $\mu$s\\
\hline
\end{tabular}
}
\caption{Complete transmission of a $s$ bytes MAC data frame at
	$d$Mbits/s.}
\label{f:transmission}
\end{figure}

We suppose that CW = 31, leading to an average backoff time of 310
$\mu$s (we indeed rarely observed a contention window larger than 31 in
our simulations, see discussions concerning the areas in
Section~\ref{s:presentation}).  Based on the previous considerations,
the waiting time $T_w$ while the neighbors are waiting corresponds to
the backoff (310~$\mu$s), the SIFS delays ($3 \times 10$~$\mu$s), the
CTS (352~$\mu$s) and ACK (304~$\mu$s) frames sent by the receiver: $T_w
= 996$.
The sending time $T_s$ while the neighbors are waiting corresponds to
the RTS (304~$\mu$s) and data frame (192 + $8s/d$~$\mu$s): $T_s = 496 +
8s/d$. Since $T_s = \alpha (T_s + T_w)$, we have
\[
  \alpha = \frac{496 + \frac{8s}{d}}{1492 + \frac{8s}{d}}
\]
In our simulations, the sending rate has been fixed to 2Mbits/s ($d=2$)
and a data MAC frame is equal to 1500 bytes ($s=1500$). We then find
$\alpha = 0.867$. This value is very close to those found in
Section~\ref{s:modeling-validation}.

\subsection{Obtaining the maximal fairness}
The previous equation shows a relationship between $\alpha$ and the
frame size $s$. We then simulated a three pairs chain while varying the
packet size. The throughput of each pair has been normalized by the
reference throughput of a single pair ($1.59$\,Mbits/s in our
configuration) in order to compute the entropy.

\begin{figure}[h!]
	\centering\includegraphics[width=\columnwidth]{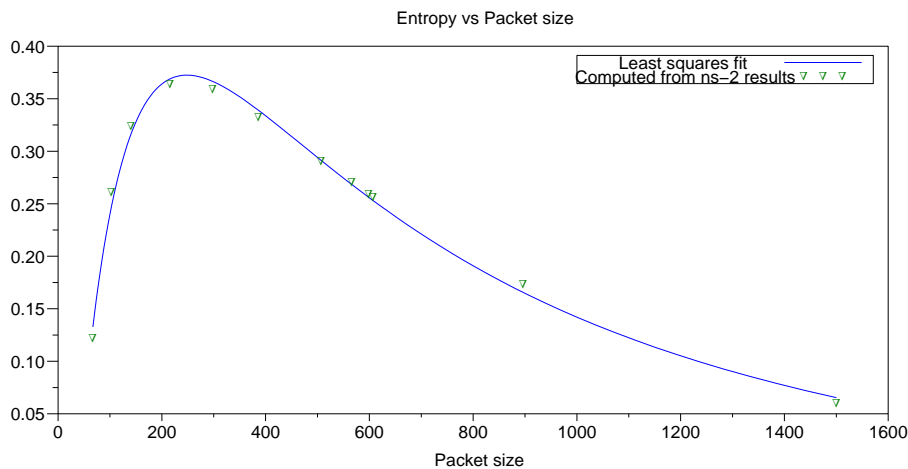}
	\caption{Entropy versus packet size.\label{f:entropy}}
\end{figure}

Results are displayed in Figure~\ref{f:entropy}. We can show that the
maximum entropy is reached for a packet size of 250 bytes. This
corresponds to $\alpha = 0.6$, which is close to the optimal $\hat
\alpha = 0.655$.

\section{Conclusion}
In this paper, we developed a scenario for ad hoc networks relying on
IEEE 802.11 wireless communications composed of a chain of senders, such
that each of them is in the carrier sense area of its neighbors. This
scenario combines the EIFS mechanism with the asymmetry of a chain,
where two nodes have only one neighbor while the others have two. This
scenario includes the three pairs fairness problem \cite{Dhoutaut02}.

We show that interesting patterns appear when the number $n$ of
sender-receiver pairs in the chain increases. These phenomena depend on
the parity of $n$. For small values of $n$, the fairness is better if
$n$ is even than if $n$ is odd. We also point out an asymptotic behavior
when $n$ increases, with a large central flat area.
By means of a simple modeling, we provide an analytical study of this
scenario, which explains the phenomena observed by simulation. Moreover,
this modeling clearly highlights a link between the fairness and the
packet size.

Besides the curious fairness phenomena we pointed out in the chain of
senders, it is interesting to notice that this simple modeling relying
on a single coefficient $\alpha$ is able to render the complex situation
of concurrent transmissions using the IEEE 802.11 standard. Previous
modeling were based on Markov chains and were not really adapted for $n$
larger than 3. This coefficient expresses the probability for a sender
to transmit a frame while its neighbors are waiting. Indeed, a sender
does not fully use the channel, even when its neighbors are waiting.

Another interesting contribution is the asymptotic results. When the
number of pairs is large, the probability of emission for a sender near
the middle of the chain is very close to the optimal value (1/3). This
optimal probability corresponds to $\alpha = 3/4$. Moreover this value
gives also the maximal fairness (expressed by means of entropy) when $n$
tends to infinity. The consequence is that, to reach the optimal case, a
sender should waste 1/4 of the time it is granted for sending. We should
also notice that when $n$ increases, the chain of IEEE 802.11
senders-receivers tends to this ideal case.

This ideal value of $\alpha$ is correct for very large values of $n$,
which does not correspond to real cases. However, for a given $n$, the
modeling is able to give the optimal $\alpha$, allowing to deduce the
(approximative) optimal packet size. When applying this method on the
chain of three pairs, we found an ideal MAC frame of 250
bytes. Simulation results with such a frame size lead to results very
close to the optimal fairness.

Among possible further works, we would like to point out other
uses of such a simple modeling, for more complex scenario.

\bibliographystyle{IEEEtran}
\bibliography{bdStrings-uk,adhoc-7}

\end{document}